\documentclass[3p,times,procedia]{elsarticle}
\usepackage{nupha_ecrc}

\volume{00}
\firstpage{1}
\journalname{Nuclear Physics A}
\runauth{}

\jid{nupha}

\jnltitlelogo{Nuclear Physics A}

\usepackage{amssymb}
\usepackage{lineno}
\usepackage{floatrow}

\usepackage[figuresright]{rotating}

\newcommand{\dbar}{$\rm\overline{d}$}

\newcommand{\s}{$\sqrt{s}$}
\newcommand{\snn}{$\sqrt{s_{\mathrm{NN}}}$}
\newcommand{\pt}{$\ensuremath{p_{\rm T}}$}
\newcommand{\dedx}{d$E$/d$x$}
\newcommand{\dndy}{d$N$/d$y$}

\newcommand{\he}{$^{3}{\mathrm{He}}$}
\newcommand{\be}{$\begin{equation}$}
\newcommand{\ee}{$\end{equation}$}

\def\muB{$\mu_B$}

\newcommand{\hypertri}{$^{3}_{\Lambda}{\mathrm{H}}$}
\newcommand{\antihypertri}{$^{3}_{\overline{\Lambda}}{\overline{\rm{H}}}$}
\newcommand{\lbd}{$\Lambda$}

\begin{document}
\begin{frontmatter}

\dochead{}

\title{Results from (anti-)(hyper-)nuclei production and searches for exotic bound
states with ALICE at the LHC}
\vspace{-0.2cm}

\author{Natasha Sharma \\
on behalf of the ALICE Collaboration}
\vspace{-0.2cm}

\address{Department of Physics and Astronomy, University of Tennessee, Knoxville, TN, USA-37996.}
\vspace{-0.2cm}

\begin{abstract}
The excellent particle identification capabilities
of the ALICE detector, 
using 
the time
projection chamber and 
the time-of-flight
detector, allow the detection of light nuclei and anti-nuclei.
Furthermore, the high tracking resolution provided by the inner tracking system 
enables the separation of primary nuclei from those coming from the decay of heavier systems. 
This allows for the reconstruction of decays such as the hypertriton mesonic weak decay \mbox{($^3_{\Lambda}$H$\rightarrow ^3$He + $\pi^-$),} the decay of a hypothetical bound state of a \lbd n into 
a deuteron and pion or the H-dibaryon decaying into a $\Lambda$, a proton and a $\pi^{-}$.
An overview of the production of stable nuclei and anti-nuclei in proton-proton, proton-lead and, in particular, lead-lead collisions is presented. 
Hypernuclei production rates in Pb--Pb are also
shown, together with the upper limits estimated on the production of hypothetical exotica candidates.
The results are compared with predictions for the production in thermal (statistical) and coalescence models.
\end{abstract}


\end{frontmatter}


\section{Introduction}
\vspace{-0.26cm}
Ultra-relativistic heavy-ion collisions at the LHC  produce a significant number 
of light (anti-)nuclei and (anti-)hypernuclei because of the large amount of energy deposited in a volume, which is much larger than in pp collisions. 
This allows to study in detail the production mechanism of nuclei and hypernuclei.
The two production mechanisms used to describe measured yields
are the coalescence model~\cite{Butler:1963pp,Kapusta:1980zz} and the statistical thermal model~\cite{Andronic:2010qu,Cleymans:2011pe}. The coalescence model is based on the simple assumption that \mbox{(anti-)}(hyper)nuclei are formed if (anti-)nucleons and/or (anti-)hyperons are close in the coordinate as well as in the momentum phase space.
In the thermal model, the abundance of particles is determined by the thermodynamic equilibrium conditions. Since the baryo-chemical potential (\muB ) is close to zero at LHC energies, the thermal model predicts the dependence of the particle yields on the chemical freeze-out temperature ($T_{\rm chem}$) by the relation \dndy\ $\propto$ $\exp(-m/T_{\rm chem})$. 
The excellent particle identification capabilities of the ALICE experiment allow for the detection of these (anti-)(hyper)nuclei and to search for exotic bound states like \lbd n and H-dibaryon. 

\vspace{-0.35cm}

\section{Data analysis}
\vspace{-0.27cm}
Nuclei and anti-nuclei such as (anti-)deuterons, (anti-)tritons, (anti-)\he\ and (anti-)$^{4}{\mathrm{He}}$ are identified 
using the specific energy loss (\dedx) measurement in the Time Projection Chamber (TPC).
 The measured energy-loss signal  of a track is required to be within a 3$\sigma$
 region around the expected value for a given particle species. 
 This method provides a pure sample of \he\ in the transverse momentum interval of 2 to 8 GeV/$c$, while it is limited up to 1.4 GeV/$c$ for deuterons. In order to extend deuteron identification, the velocity measurement by the  Time-Of-Flight detector (TOF) is used.  
The momentum range under study is only limited by the available statistics and not by the detector performance.
 The secondaries due to knock-out from the detector material are rejected by applying a cut on the Distance-of-Closest Approach along the beam axis, $|$DCA$_Z|$  $<$ 1.0 cm. 
 This selection removes a large fraction of 
 background for nuclei, but does not affect primary anti-nuclei. 
The measured raw spectra are then corrected for efficiency and detector acceptance. More details can be found in Ref.~\cite{Adam:2015vda} and references therein.

The production of (anti-)hypertriton, \hypertri , \antihypertri\ has been measured in Pb--Pb collisions via the invariant-mass reconstruction of the mesonic weak decay
channel \hypertri\ $\rightarrow$ \he\  + $\pi^{-}$ and \antihypertri\ $\rightarrow$ $^{3}{\overline{\rm{He}}}$ + $\pi^{+}$, respectively.
Topological selections are applied in order to identify secondary decay vertex and
to reduce the combinatorial background.
More details about the used analysis technique can be found in Ref.~\cite{Adam:2015yta}.

\vspace{-0.35cm}
\section{Results and discussions} 
\vspace{-0.2cm}
\subsection{Nuclei and anti-nuclei mass difference}
\begin{figure}
\vspace{-0.2cm}
\floatbox[{\capbeside\thisfloatsetup{capbesideposition={right,center},capbesidewidth=5cm}}]{figure}[\FBwidth]
{\caption{Mass-over-charge ratio difference for d-\dbar\ and \he-$^{3}{\overline{\rm{He}}}$ compared with the CPT invariance expectation (dotted lines). 
The solid red points show the ALICE measurements and the open black circles show the existing mass  difference measurements.
Error bars represent the quadrature sum of the statistical and systematic uncertainties (standard deviations)~\cite{Adam:2015pna}.}\label{massDiff}}
{\includegraphics[width=4.7cm]{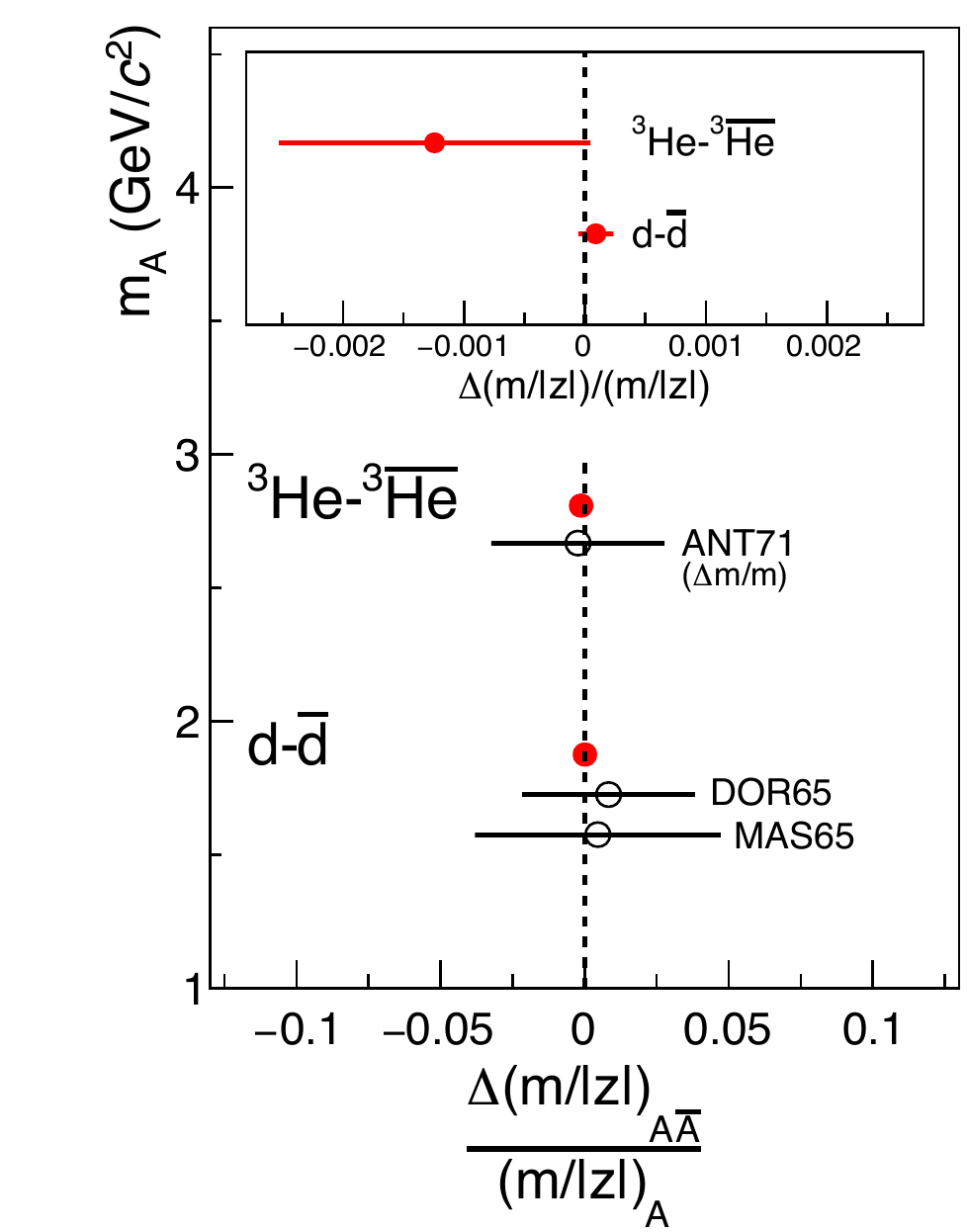}}
\vspace{-0.7cm}
\end{figure}

\vspace{-0.1cm}
The momentum-over-charge ($p/z$) measurement from the TPC and velocity measurement using the TOF allow to obtain mass-over-charge distributions for (anti-)deuterons and (anti-)\he. 
Figure~\ref{massDiff} shows the ALICE measurements of the mass-over-charge ratio difference for d-\dbar\ and \he-$^{3}{\overline{\rm{He}}}$. 
The results are compared with the CPT invariance expectation and with the existing mass measurements. 
These measurements show that the mass and binding energies of nuclei and anti-nuclei are compatible within uncertainties, confirming CPT  invariance for light nuclei~\cite{Adam:2015pna}.

\vspace{-0.25cm}
\subsection{Transverse momentum distributions, ratios, and yields}
\begin{figure}
\vspace{-0.15cm}
  \includegraphics[width=7.3cm]{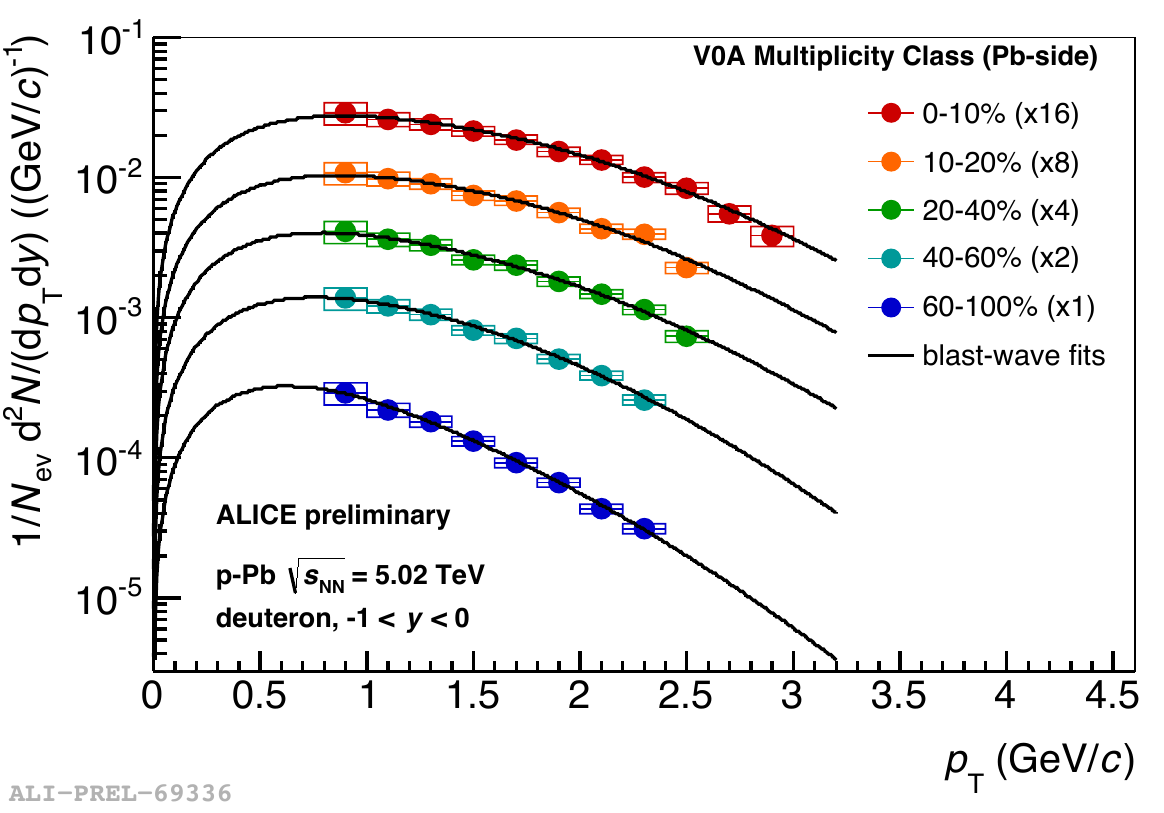}
  \includegraphics[width=6.3cm]{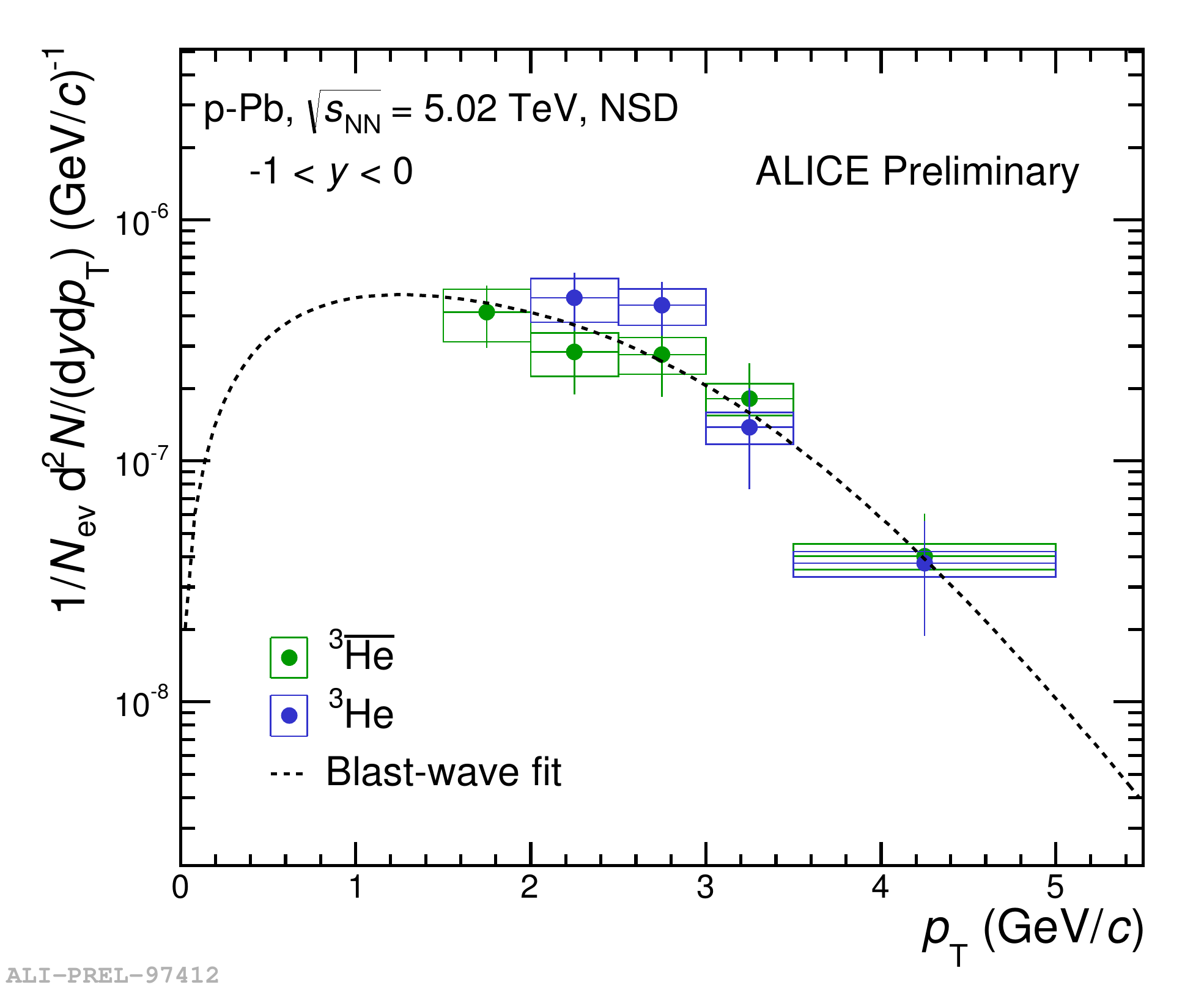}
 \caption{The transverse momentum distribution of deuterons (left panel) and of \he\ (right panel) for p--Pb collisions at \snn\ = \mbox{5.02 TeV.} The boxes indicate systematic uncertainties, whereas the lines represent statistical uncertainties.}
\label{spectra}
\vspace{-0.25cm}
\end{figure}

\vspace{-0.1cm}
The deuteron transverse momentum (\pt) distributions are obtained for Pb--Pb collisions at \snn\ = 2.76 TeV~\cite{Adam:2015vda},
and for p--Pb collisions at \snn\ = 5.02 TeV for various centrality classes, and also for pp collisions at \s\ = 7 TeV~\cite{Adam:2015vda}.
A hardening of the spectrum with increasing centrality is observed in both Pb--Pb and p--Pb collisions.
The \pt\ distributions of \he\ are obtained for two centrality classes in Pb--Pb collisions at \snn = 2.76 TeV~\cite{Adam:2015vda}
and for non-single diffractive (NSD) p--Pb collisions at \snn = 5.02 TeV. As an example, Fig.~\ref{spectra} shows the \pt\ distribution of deuterons and \he\ in p--Pb collisions.
The V0A multiplicity classes (Pb-side) corresponds to the measurement of the multiplicity in the Pb-going direction using the V0A detector.
In order to extrapolate the yield into the unmeasured \pt\ region, the spectra are fitted individually with a Blast-Wave function. 

\begin{figure}[h]
\vspace{-0.3cm}
 \begin{center}
  \includegraphics[width=8.5cm]{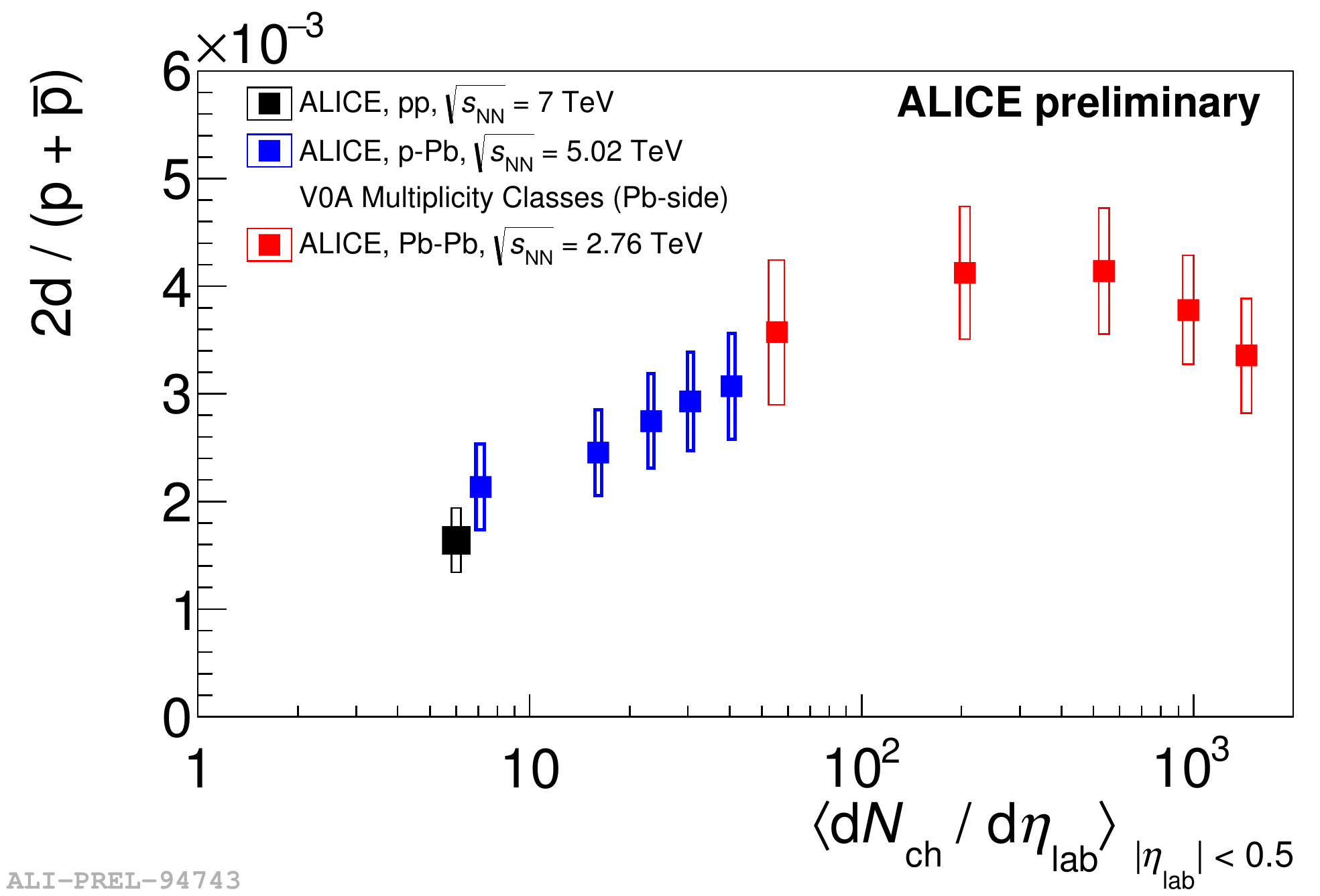}
 \caption{d/p ratio as a function of charged particle multiplicity for different colliding systems at LHC energies.}
\label{D2p}
 \end{center}
 \vspace{-0.6cm}
\end{figure}

If nuclei production is described by the thermal model then the d/p ratio is expected to remain constant for all colliding systems and for all multiplicities (because \muB$\approx$0).
Figure~\ref{D2p} shows the ratio of deuteron-to-proton yields as a function of event multiplicity for pp, p--Pb, and Pb--Pb collisions. 
Within the uncertainties no significant variation with multiplicity is observed in Pb--Pb, which is consistent with the thermal 
model expectations~\cite{Adam:2015vda}. 
In p--Pb, the d/p ratio increases with multiplicity, which is incompatible with the thermal model expectations for chemical freeze-out temperatures that do not depend on multiplicity.
The ratio in pp collisions is a factor of 2.2 lower than in Pb-Pb.

\begin{figure}
\vspace{-0.15cm}
 \includegraphics[width=6.5cm]{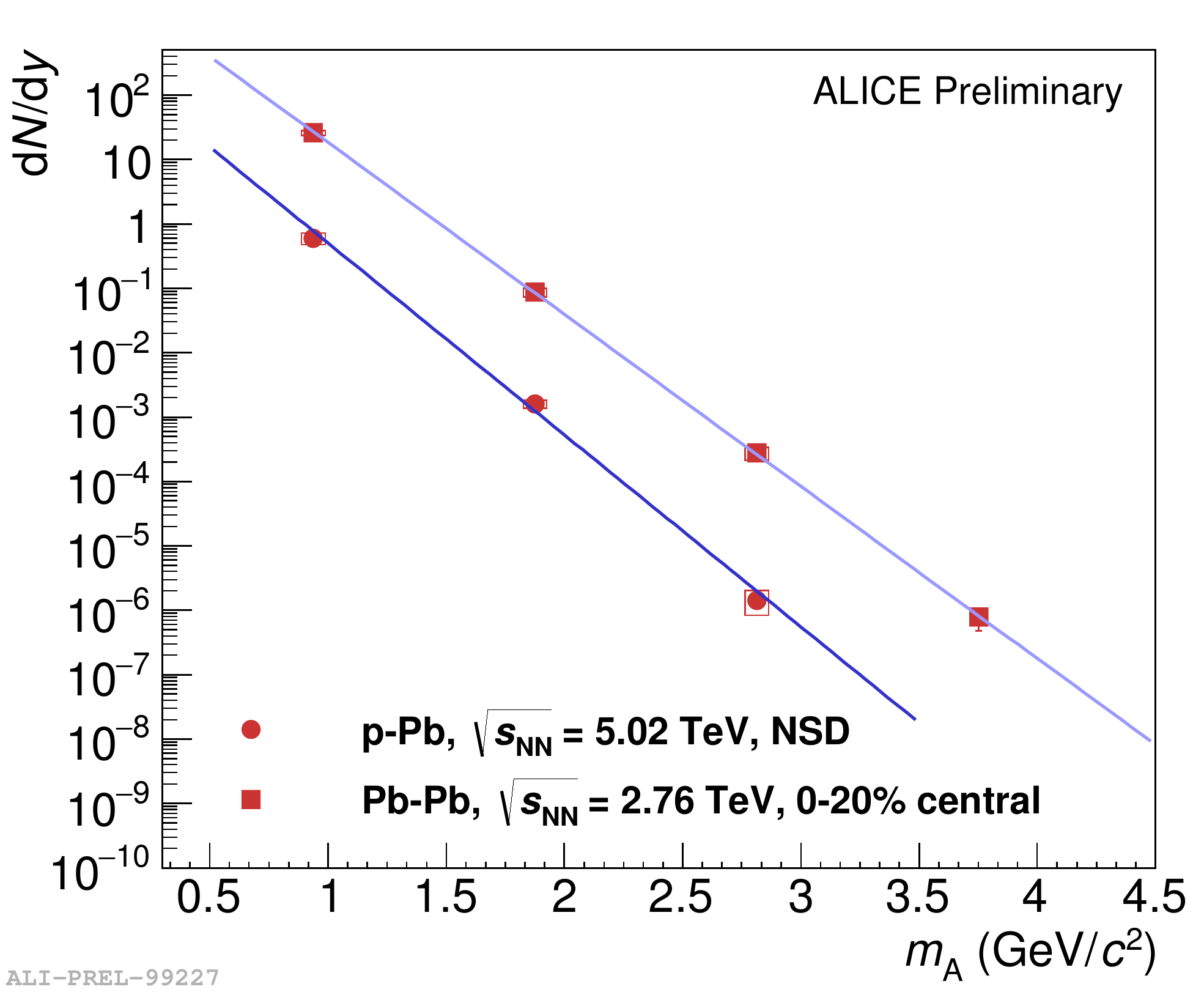}
    \includegraphics[width=7.5cm]{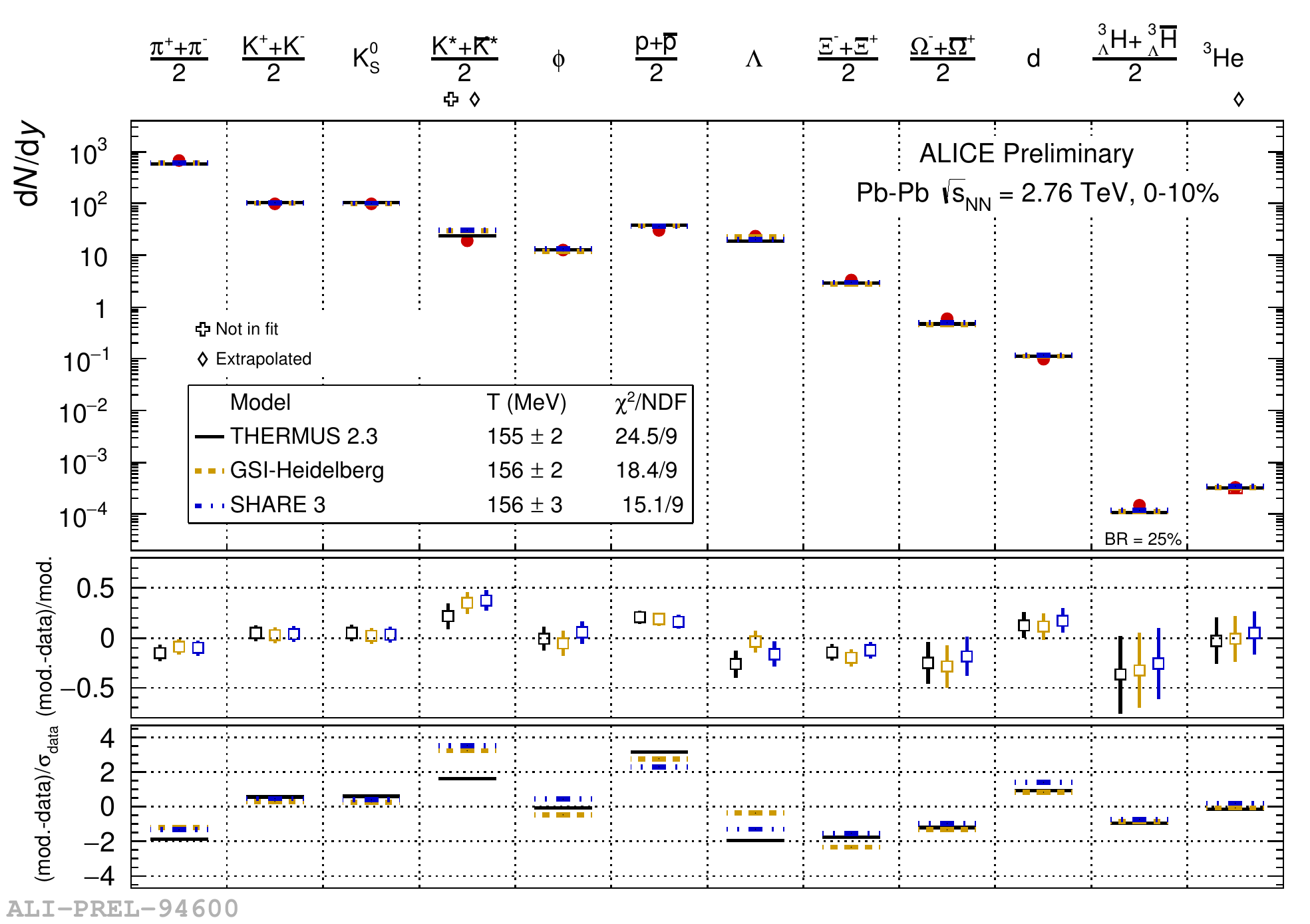}
 \caption{Left panel: The production yield \dndy\  of light nuclei as a
function of the particle mass $m_{{\rm A}}$ measured for 0-20\%
central Pb--Pb collisions at \snn\ = 2.76 TeV and for NSD p--Pb collisions at \snn\ = 5.02 TeV. The boxes indicate systematic uncertainties, whereas the lines represent statistical uncertainties
The lines represent fits with an exponential function.
 Right panel: Thermal model fit to various particle yields for 0-10\% central Pb--Pb collisions at \snn\ = 2.76 TeV. }
\label{thermalModel}
\vspace{-0.25cm}
\end{figure}

The left panel of Fig.~\ref{thermalModel} shows the mass dependence of light nuclei yield for central Pb--Pb collisions and for NSD p--Pb collisions. 
The lines represent a fit with an exponential function. The figure also shows the anti-alpha yield measured in Pb--Pb collisions. Nuclei yields follow an exponential decrease with the mass.   The penalty factor,
namely the reduction of the yield by adding one nucleon, is $\sim$300 for Pb--Pb collisions and is $\sim$600 for NSD p--Pb collisions. 
The thermal model predicts the exponential dependence of the yield on the temperature given by \dndy\ $\propto$ $\exp(-m/T_{\rm chem})$.
The slope of the exponential fits can be used to extract $T_{\rm chem}$. 
For Pb--Pb collisions, the obtained value is consistent with the thermal model expectation. 
However, the corresponding value is much lower for p--Pb collisions. 
Figure~\ref{thermalModel} (right) shows the thermal model fit to various particle yields including light (hyper-)nuclei for 0-10\% central Pb--Pb collisions at \snn\ = 2.76 TeV. 
Different models 
like~\cite{Andronic:2010qu,Wheaton:2004qb,Torrieri:2006xi} 
describe particle yields including deuterons, \he\ and \hypertri\  yield well using $T_{\rm chem}$$\sim$156 MeV.
Exclusion of the nuclei from the fit does not cause any significant change in $T_{\rm chem}$.

\vspace{-0.5cm}
\subsection{Searches for exotica}
\begin{figure}[h]
\begin{center}
  \includegraphics[width=11.0cm]{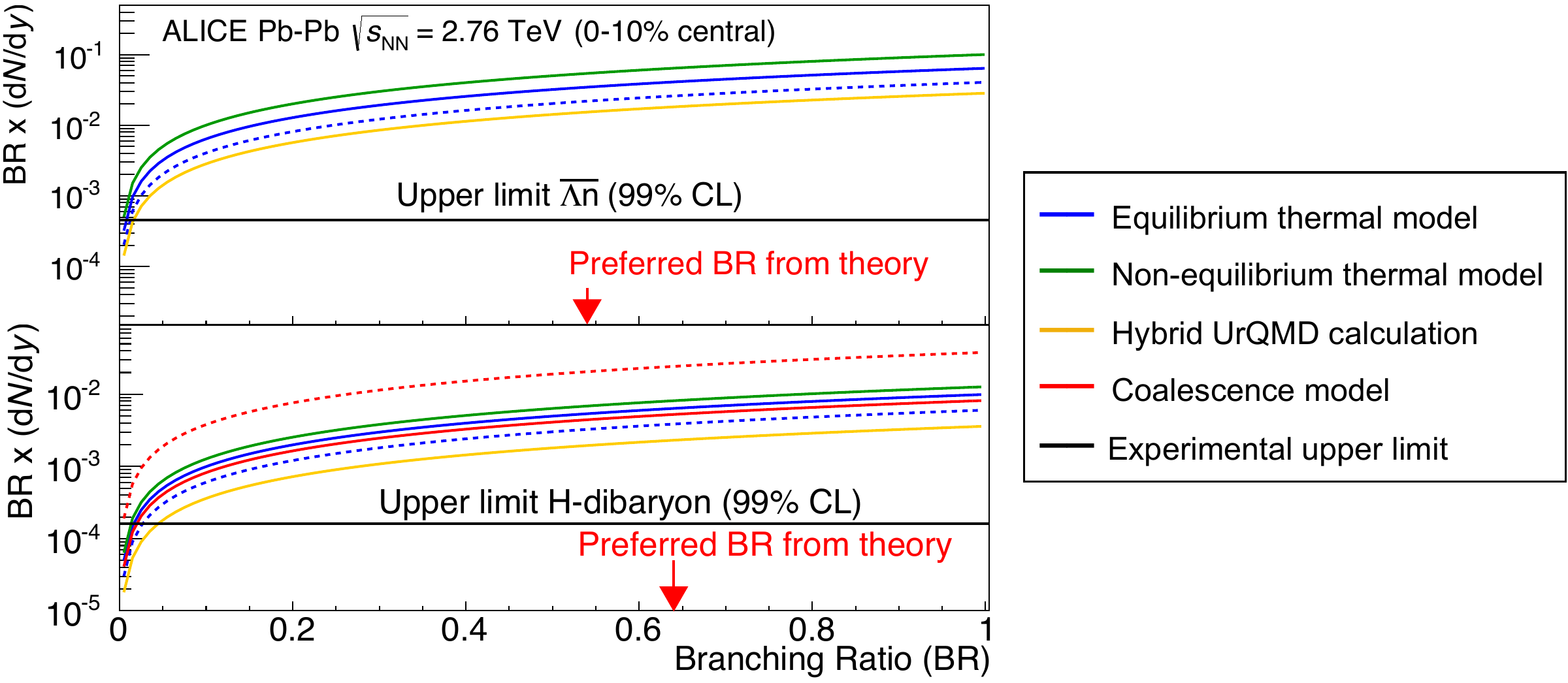}
 \caption{ Experimentally determined upper limit, under the assumption of the lifetime of a free \lbd. In the upper panel shown for the \lbd n bound state and for the H-dibaryon in the lower panel. The theory lines are drawn for different branching ratios (BR)~\cite{Adam:2015nca}.}
\label{ExoticaModelComp}
 \end{center}
 \vspace{-0.6cm}
\end{figure}

The thermal model describes the (hyper)nuclei yields well for Pb--Pb collisions and therefore can be used for predicting 
the expected yields of 
weakly decaying exotic bound states like H-dibaryon (\lbd\lbd) and $\overline{\rm \Lambda n}$. 
The possible existence of these weakly bound states 
has been investigated via the decay channel \lbd\lbd $\rightarrow$  \lbd + p + $\pi^-$ and $\overline{\rm \Lambda n}$ $\rightarrow$ $\bar{d}$ + $\pi^+$. 
No evidence has been seen in the invariant-mass distribution, neither for the \lbd\lbd\ nor for the $\overline{\rm \Lambda n}$  bound state.
The upper limits on the yield  of \lbd\lbd\ and $\overline{\rm \Lambda n}$ are about a  factor 20 lower than the thermal model predictions
when assuming reasonable lifetimes and branching ratios (BR).
Figure~\ref{ExoticaModelComp} shows the experimentally determined upper limit for \lbd\lbd\ and $\overline{\rm \Lambda n}$ bound states compared with the model calculation as a function of BR, see Ref.~\cite{Adam:2015nca} for more details.

\vspace{-0.7cm}
\section{Summary and conclusions}
\vspace{-0.2cm}
The (anti-)nuclei production has been measured by the ALICE Collaboration in pp, p--Pb, and Pb--Pb collisions. 
Hardening of deuteron spectra with increasing centrality is observed in p--Pb and Pb--Pb. 
The d/p ratio rises with multiplicity in p--Pb and remains almost constant in Pb--Pb.
The nuclei yields follow an exponential decrease with mass. 
This exponential decrease in Pb--Pb reflects a temperature similar to $T_{\rm chem}$ expected from the thermal fits of various produced particles suggesting that the nuclei follow the thermal behavior. However, in p--Pb the obtained temperature is less than the expected $T_{\rm chem}$ suggesting nuclei might not follow thermal behavior in p--Pb.
Statistical hadronization models such as THERMUS~\cite{Wheaton:2004qb}, SHARE~\cite{Torrieri:2006xi}
describe
particle and light (hyper)nuclei yields 
well at $T_{\rm chem}$$\approx$156 MeV for Pb--Pb. 
The upper limits for \lbd \lbd\ and $\overline{\rm \Lambda n}$ are lower than the thermal model expectations by at least an order of magnitude. Therefore the existence of such states with the 
BRs, masses and lifetimes typically assumed in literature is questionable.


\vspace{-0.25cm}


\begin{thebibliography}{9}

\vspace{-0.25cm}
\bibitem{Butler:1963pp}
  S.~T.~Butler and C.~A.~Pearson,
  Phys.\ Rev.\ Lett.\  {\bf 7} (1961) 69.
  
  \bibitem{Kapusta:1980zz} 
  J.~I.~Kapusta,
  Phys.\ Rev.\ C {\bf 21} (1980) 1301.

\bibitem{Andronic:2010qu} 
  A.~Andronic, P.~Braun-Munzinger, J.~Stachel and H.~St\"{o}cker,
  Phys.\ Lett.\ B {\bf 697} (2011)  203.

\bibitem{Cleymans:2011pe} 
  J.~Cleymans, S.~Kabana, I.~Kraus, H.~Oeschler, K.~Redlich and N.~Sharma,
  Phys.\ Rev.\ C {\bf 84} (2011) 054916.
  
  
\bibitem{Adam:2015vda} 
  J.~Adam {\it et al.} [ALICE Collaboration],
  arXiv:1506.08951 [nucl-ex].
  
  
  
\bibitem{Adam:2015yta} 
  J.~Adam {\it et al.} [ALICE Collaboration],
  arXiv:1506.08453 [nucl-ex].

\bibitem{Adam:2015pna} 
  J.~Adam {\it et al.} [ALICE Collaboration],
  Nature Phys.\  {\bf 11} (2015) 811.

\bibitem{Wheaton:2004qb} 
  S.~Wheaton and J.~Cleymans,
  Comput.\ Phys.\ Commun.\  {\bf 180} (2009) 84.



\bibitem{Torrieri:2006xi} 
  G.~Torrieri, S.~Jeon, J.~Letessier and J.~Rafelski,
  Comput.\ Phys.\ Commun.\  {\bf 175} (2006) 635.

\bibitem{Adam:2015nca} 
  J.~Adam {\it et al.} [ALICE Collaboration],
  Phys.\ Lett.\ B {\bf 752}  (2016) 267.

  
  
  
\end{thebibliography}
\end{document}